\documentclass[pop,12pt,fleqn]{w-art}
\usepackage{times}
\usepackage{w-thm}

\numberwithin{equation}{section}
\begin{document}
\DOIsuffix{theDOIsuffix}
\Volume{51}
\Issue{1}
\Month{01}
\Year{2003}
\pagespan{1}{}



\baselineskip=14pt
\parskip=3pt


\bigskip

\begin{flushright}
\hfill{LMU-ASC 81/05}
 \\
\hfill{hep-th/0601088}
~ 
\end{flushright}

\title[Black Holes and Rings]{Black Holes and Rings of
minimal 5-dim.\ Supergravity\footnote{To appear in Proceedings of: 
RTN workshop: ``The quantum structure of space-time and the geometric 
nature of fundamental interactions''
(Corfu, September 2005)
}
}
 

\author[Klaus Behrndt]{Klaus Behrndt}
 \address[]{ \em
Arnold-Sommerfeld-Center for Theoretical Physics \\
Department f\"ur Physik, Ludwig-Maximilians-Universit\"at 
M\"unchen \\
Theresienstra\ss{}e 37, 80333 M\"unchen, Germany \\[1mm]
\rm and \\[1mm]
\em
Max-Planck-Institut f\"ur Physik,
F\"ohringer Ring 6, 80805 M\"unchen, Germany \\[3mm]
E-mail: {\sf behrndt@theorie.physik.uni-muenchen.de}
}


\bigskip

\begin{abstract}
  In this talk, I will summarize recent developments in 5-dimensional
  supergravity. Apart from black ring solutions, we will discuss the
  way of obtaining regular (bubbling) solutions with the same charges
  as black holes. We outline the procedure for the solution in five
  and four dimensions. Finally we explore the close relationship
  between 4- and 5-dimensional supersymmetric stationary solutions.
\end{abstract}


\maketitle                   






\newcommand{\be}[3]{\begin{equation}  \label{#1#2#3}}
\newcommand{\ee}{\end{equation}}
\newcommand{\ba}{\begin{array}}
\newcommand{\ea}{\end{array}}
\newcommand{\bea}[3]{\begin{eqnarray}  \label{#1#2#3}}
\newcommand{\eea}{\end{eqnarray}}
\newcommand{\com}[1]{\begin{center} {\tt #1} \end{center}}
\newcommand{\ip}{\raise1pt\hbox{\large$\lrcorner$}\,}

\newcommand{\N}{${\cal N}\, $}

\def\I{\mathbf I}
\def\P{\mathbf P}
\def\R{\mathbf R}
\def\S{\mathbf S}
\def\X{\mathbf X}
\def\C{\mathbf C}
\def\Y{\mathbf Y}
\def\1{\mathbf 1}

\newcommand{\haken}{\mathbin{\hbox to 8pt{%
                 \vrule height0.4pt width7pt depth0pt \kern-.4pt
                 \vrule height4pt width0.4pt depth0pt\hss}}}



\section{Introduction}


The aim of this talk is to address recent progress of 5-dimensional
supergravity.  We start with an issue related to the uniqueness theorem of
black holes, which states that the most general 4-dim.\ black hole has to be
of the Kerr-Newmann type, ie.\ is uniquely fixed by its conserved charges
(mass, angular momentum, electric/magnetic charges) and the horizon topology
is a 2-sphere \cite{Isra,Ha72}. It implies especially that any wormhole has to
be hidden behind a spherical horizon (topological censorship) and hence
appears as a black hole for an outside observer.  It fueled moreover the idea
that stable microscopic black holes can regarded as elementary particles
\cite{9202014}, which however is not (yet) realistic for different reasons,
eg.\ there are no stable spinning black holes in four dimensions and moreover
the mass to charge ratio is either not realistic or moduli dependent, but one
should re-address this issue once all moduli are fixed. On the other hand, one
can also ask, what happens in higher dimensions? As it turns out, the topology
can be different. In five dimensions for examples, black rings with the
geometry $S_1 \times S_2$ have been found \cite{0110260,0407065} and for
$D>5$, the allowed topology might be even more rich. This is in agreement with
the general statement \cite{9902061}, that the horizon manifold has to have
positive scalar curvature if the asymptotic vacuum is flat, which forbids in
four dimensions, for example, a $T^2$ topology of a black hole horizon.

The second issue that I want to address in my talk is related to the
information loss due to black holes.  Any information about states passing an
event horizon seems to be lost -- one can recover only thermal radiation and
this violates obviously the unitary evolution of quantum mechanical processes.
The corresponding calculation has been done so far in a semiclassical
approximation and one expects that a consistent theory of quantum gravity will
eventually resolve this problem. In string theory encouraging results are
obtained, which however rely mostly on supersymmetry or assume a weakly
coupled regime, where black holes can be approximated by an intersection of
branes embedded in flat space time \cite{9601029, 9611140}. New light on this
issue is shed by the conjecture of Lunin and Mathur, who propose to
``resolve'' black holes into a smooth geometry carrying the same conserved
charges \cite{0105136,0502050}. Following this conjecture, black holes are
fuzzballs of extended string states and the appearance of an event horizon is
only an artifact of the parameterization and there is always an equivalent
bound state with a regular geometry. As we will see below, for a large class
of supersymmetric black holes, this is in fact possible. But it is unclear in
the moment whether something analogous exists also for the Schwarzschild
solution, for example.

Finally, we comment on the mapping of 4- and 5-dim.\ supersymmetric black
holes. This mapping is interesting because it opens the possibility to
estimate unknown corrections to the 5-dim.\ supergravity from known corrections
in four dimensions. These are, for example, higher derivative couplings (eg.\
higher curvature terms), which are understood in four \cite{9812082, 9801081}
but not in five dimensions. Moreover, the entropy of supersymmetric 4-dim.\
black holes is the Legendre transform of the free energy, which can be
calculated in topological string theory \cite{0405146}. One may wonder about
the analog statement in five dimensions, which will be related to the M-theory
limit of the topological string theory.  In fact, the one-to-one map relates
{\em any} supersymmetric stationary solution of 4-d supergravity to a {\em
localized} solution in five dimensions with electric charges and non-vanishing
magnetic dipole charges and  implies especially that the corresponding
(Bekenstein-Hawking) entropies coincide.


\section{Classifying supersymmetric solutions of minimal supergravity}


Before we discuss the relation of 4- and 5-dim.\ supergravity and the
resolution of singularities, we should summarize what is known about
supersymmetric solutions in five dimensions \cite{0209114}.  In contrast to
four dimensions, 5-dim.\ black holes have in general two angular momenta $J_1$
and $J_2$ and the prime example of supersymmetric black holes in five
dimensions is the BMPV solution \cite{9602065}, where both angular momenta are
equal and which is charged under an $U(1)$ gauge group. This solution has been
generalized by coupling it to additional $U(1)$ gauge groups and scalar fields
\cite{9603100} or to general \N = 2 vector multiplets in
\cite{9708103,9801161}. It is straightforward to construct also a magnetic BPS
string solution, which is infinitely extended and lifts to intersecting M5
branes in 11 dimensions. All these solutions rely however on certain
assumptions for the geometry, the charges and/or the choice of harmonic
functions, because they were not derived in complete generality.  For minimal
supergravity (ie.\ with 8 supercharges or \N = 2), this analysis was done by
Gauntlett etal.\ \cite{0209114} by redoing basically the analysis of Tod in
four dimensions \cite{tod}. The basic idea relies only on supersymmetry, which
requires the existence of Killing spinors $\epsilon^a$ ($a =1,2$) which can be
combined into one scalar $f = \varepsilon_{ab} \bar \epsilon^a \epsilon^b$ ,
one vector $V_\alpha = \varepsilon_{ab} \bar \epsilon^a \gamma_\alpha
\epsilon^b$ , and three 2-forms $\Phi^{ab}_{\alpha\beta} = \bar \epsilon^a
\gamma_{\alpha\beta}\epsilon^b$ (symmetric in $a,b$). In the absence of
sources, these spinors are globally well-defined and thus also the fermionic
bi-linears are globally well-defined and, due to Fierz identities, the vector
$V$ is timelike with $|V|^2 = - f^2$ or null. The Killing spinor equations
imply that the 2-forms are closed and have no components along $V$, which is a
Killing vector. For the null case, it is pp-wave type solution (which we will
not discuss here) and if $V$ is timelike and we have stationary solutions,
with hyper K\"ahler base space. Therefore the metric can be written as
\be821
ds^2 = -f^2 (dt + \omega)^2 + f^{-1} h_{mn} dx^m dx^n
\ee
where $h_{mn}$ is the hyper K\"ahler metric on the base space. By
employing the Killing spinor equations, one finds moreover
\cite{0209114}
\be911
F = de^0 - G^{+} \quad , \quad
f d\omega = G^{+} + G^{-} \quad , \quad \Delta f^{-1} = (G^{+})^2 
\ee
where $G^{\pm}$ are (anti) selfdual forms on the base and $dF =0$
implies that $G^{+}$ has to be closed. Now, a given solution is
related to a choice of these forms and $\omega$ and $f$ are solutions
to these equations.  For example, a static metric is related to
$G^{-}=0$ or if the base space is compact and smooth, one obtains
$f=1$, $G^{+} = 0$ whereas for the non-compact case $f=1$ and $G^{+} =
0$ yields the G\"odel type solution.  The BMPV black hole corresponds
to $G^{+} = 0$ and a flat base space. In general, the base space can
be any hyper K\"ahler space, but especially interesting is a
Gibbons-Hawking space \cite{GiHa}, which allows for a tri-holomorphic
Killing vector and the corresponding dimensional reduction yields the
known 4-dim.\  (stationary) solutions. Denoting the Killing
vector by $\partial_\psi$, this metric can be written as
\be729
h_{mn} dx^m dx^n = \frac{1}{N} ( d\psi + A)^2 + N d \vec x d \vec x
\quad , \qquad {^{\star_3}}dA = dN 
\ee
and thus $N$ is 3-d harmonic function in $\vec x$. For the simple case $N=
1/|\vec x|$ it is nothing but flat space and for $N= 1 + n/|{\vec x}|$ it is a
Taub-NUT space with the NUT charge $n$. The general solution to the eqs.\
(\ref{911}) becomes
\bea613
\omega &=& \omega_5 (d\psi + A) + \vec \omega 
\quad , \qquad \omega_5 = M + \frac{L K}{N} +
\frac{K^3}{N^2} \quad , \qquad
f^{-1} = L + \frac{K^2}{N}  \label{943}
\eea
where $(N, K , L, M)$ are harmonic functions are their poles are
related to the following conserved charges: $M \sim $ angular
momentum; $N \sim $ NUT charge; $L \sim $ electric charge ($Q \sim
\int_{S_3} {^\star}F$); $K \sim $ magnetic dipole ($P \sim \int_{S_2}
G^{+}$).

In order to get the black ring solution with the horizon topology of
$S_1 \times S_2$, one has to consider a 2-center solution. For
example, a single black ring in a flat space at $\vec x = \vec x_0
=(0,0,-a)$ \cite{0407065,0408010} requires
\bea820
&&  N = 1/r \quad , \qquad
K = p/\Sigma \quad , \qquad L = 1+q/\Sigma \quad , \qquad
M = -p(1 - a/\Sigma) \quad ,  \\
&& \vec \omega  = p (\cos\theta + 1)(1-\frac{a +r}{\Sigma}) d\phi 
\quad , \qquad
\Sigma = |\vec x - \vec x_0|= \sqrt{r^2 + a^2 + 2 ar \cos\theta}
\eea
with $r = |\vec x|$ and the three parameters $p, q, a$ correspond to the
magnetic dipole charge, the electric charge and the rotational
parameter. Lifting this solution to 10 dimensions, yields the supertube
solution of Emparan, Mateos and Townsend \cite{0106012}, which comprises a
fundamental string, a D0-brane and the rotation generates a D2 dipole
charge. We should also mention the non-extremal black ring solution has been
found, see \cite{0412130} for the most general example.


\section{Bubbling solutions}


Lunin and Mathur constructed a solution with the same conserved
charges as a black hole, but without a singularity or horizon
\cite{0105136}. This ``fuzzball'' expands into a certain volume
bounded by a surface whose area matches the Bekenstein-Hawking
entropy. This surface grows with the degeneracy and the notion of an
entropy is introduced by a coarse graining process, where one traces
over the states within the volume. Moreover, they conjecture that
string theory can always resolve (at least supersymmetric) black holes
in fuzzballs of string states. In the strong version of this
conjecture, also non-supersymmetric black holes as the Schwarzschild
solution, for example, are expected to grow into fuzzballs. If this
conjecture is true, there is no information paradox since all states
remain visible.

But how is this possible? How can a horizon disappear? To understand
the answer, we have to recall that in compactified string/M-theory the
horizon of a supersymmetric black hole is the location of
(intersecting) branes. Especially 5-dim.\ black holes appear as
intersecting M2-branes where the 4-form field strength solves the
11-dim.\ equation of motion
\be822
d^{\star}F_4 = - Q_{M2} \, \delta^{(8)} + F \wedge F \ .
\ee
After Calabi-Yau compactification this equation becomes
\be722
\Delta f^{-1} = - Q \, \delta^{(4)} + (G^{+})^2
\ee
where the selfdual 2-form $G^{+}$ is the one introduced in the last section.
Expanding it in a set of harmonic 2-forms, the solution is given by
(\ref{943}) and the regular (bubbling) solution corresponds to the case, where
the different sources of the harmonic functions cancel against each
other. This means, that the M2-brane charge entering the function $L$ is
canceled by the dipole charges in $K$ (related $G^{+}$) so that $f^{-1}$
becomes constant near a given center.  {From} the solution of $f$ in
(\ref{943}), we see that this is only possible if $N$ has also a pole at that
center, ie.\ the NUT fiber has to degenerate at that point. Note, the dipole
charges come from wrapped M5-branes and two intersecting M5-branes carry an
effective membrane charge due to the $F \wedge F$ term in (\ref{822}).  {From}
the microscopic point of view, this membrane charge is due to a small
instantons of the self-dual tensor field on the five brane world volume.  In
simple supergravity with only one $U(1)$ gauge group this cancellation would
imply that $f$ becomes constant and the solution is trivial. We get only a
non-trivial solution if the gauge group is $[U(1)]^n$ with $n>1$ and in
addition, if the solution has different centers so that at each center, a
different charge is canceled.  The modifications due couplings to additional
vector multiplets have been derived in \cite{0408122}.  The selfdual 2-form
$G^{+}$ has to be replaced by $X_A \Theta^A$, where $A=1 ...  n$ counts the
different vector multiplets with the gauge fields
\be511
F^A = d(X^A e^0 ) + \Theta^A \ .
\ee
Apart from gauge fields, each vector multiplet of 5-dim.\ minimal
supergravity comes with a real scalar and the corresponding moduli
space is defined by the cubic equation: ${\cal V} \equiv \frac{1}{6}
C_{ABC}X^A X^B X^C = 1$ with real $X^A$ and $C_{ABC}$ denoting the
topological intersection numbers.  The metric can then be written as
\bea166
ds^2 = - f^2 (dt + \omega)^2 + f^{-1} h_{mn} dx^m dx^n 
&,&  \omega = \omega_5 (d\psi + A) + \vec \omega d\vec x \ , \\
\omega_5 = M + \frac{L_A K^A}{N} +  
\frac{C_{ABC} K^A K^B K^C}{3\, N^2} &  , &
{^\star}d \vec \omega = N dM -M dN + K^A dL_A - L_A dK^A \ . \nonumber
\eea
The value of the function $f$ and the scalar fields are now
fixed by the algebraic equations
\be766
\frac{1}{2} C_{ABC} Y^B Y^C = e^{2/3} 
\Big(L_A + \frac{C_{ABC} K^B K^C}{2 \,N} \Big)
\equiv \Delta_A \qquad {\rm with:} \quad Y^A = X^A/\sqrt{f}
\ee
and the solution is fixed by the set of harmonic functions $M$, $N$,
$L_A$ and $K^A$, ie.\ for each vector multiplet we have
two functions or two charges (one electric charge and one magnetic
dipole charge). The remaining two charges correspond to the graviphoton
which enters the gravity multiplet. 

The procedure to obtain a regular (bubbling) solution is now as
follows \cite{0505166,0505167}.  One takes a multi-center solution
(eg.\ for each $U(1)$ a separate center) and cancels the poles in the
harmonic function at each center so that
\be726
\Delta_A \Big|_{x= x_i} < \infty \qquad {\rm and}\qquad
\omega_5\Big|_{x = x_i} < \infty \ .
\ee
There are two ways to understand these equations, either they fix the
dipole charges (entering $K^A$) or one sees these equations as a
fixing for moduli, which are related to the constant parts in the
harmonic functions, see below. But these constraints are not yet
enough for a well-defined supergravity solution. One has also to proof
that there are no closed time curves or Misner strings, which requires
\be112
d\vec \omega = 0
\ee
at each center. This will give a constraint on the position of the
centers, which means that the moduli related to the positions in the
external space are also lifted. We should also mention, that
this procedure involves a subtle point. Namely, the cancellation
requires some negative poles. For example in the simplest case
one has to take as $N$ function
\be711
N = \frac{1}{r} \, + \, Q\,  \Big( \frac{1}{r_i} - \frac{1}{r_j}\Big) + ...
\ee
where the first term ensures that the base space is asymptotically
flat.  At the position $\vec x = \vec x_j$ is now a negative pole,
which implies that the hyper K\"ahler base space changes its signature
when $N$ changes its sign and moreover it becomes singular at zeros of
$N$. So, the base space is not well-defined everywhere, but the
5-dim.\ solution remains smooth!  The simplest way to see this, is to
investigate the corresponding 4-dim.\ solution, which is smooth at
this point. But if the conditions (\ref{726}) are satisfied, the
4-dim.\ solution is singular at the poles of $N$ [see eq.\
(\ref{299})] and is regular only if $N=const$. To make this more
clear, lets discuss an example.

\medskip

{\em Bubbling STU model}

\noindent
The STU-model is defined by ${\cal V} = STU =1$ and we have
\bea721
\Delta_1 &=& L_1 + 2 \, \frac{K^2 K^3}{N} \ , \\
\Delta_2 &=& L_2 + 2 \, \frac{K^1 K^3}{N} \ , \label{552} \\
\Delta_3 &=& L_3 + 2 \, \frac{K^1 K^2}{N} \ . \label{233}
\eea
An interesting question is, what is the least number of centers for
which a bubbling solution exists? Before we discuss this, let us
comment on the moduli. The complete solution is fixed by the set of
harmonic functions and their constant parts as well as the different
positions of the centers appear as moduli.  If $N \neq 0$ (yielding a
singular 4-dim.\ solution), we need only two centers and it is a
straightforward exercise to distribute the harmonic functions so that
the relations (\ref{726}) are obeyed. For the STU model each center
gives four equations and hence we fix $2 \times (3+1)$ constants,
e.g.\ the constant parts in the harmonic functions. If we do not want
to impose constraints on the charges, the integrability constraint in
(\ref{413}) yields another condition on the relative position; see
\cite{0005049,0506251}. Note, even if we want to set $\vec \omega =0$,
this constraint applies.

But how can we obtain a regular solution in four dimensions?  This is
the case, if we impose $N=const.$, which in turn implies that
$A=consts$ in (\ref{729}) and hence the hyper K\"ahler base becomes
$S_1 \otimes R_3$. Now, to cancel the poles in $\Delta_A$, we have to
distribute the poles in $K^A$ at different centers. For the STU-model
this is done if we assume three centers, so that each $K^i$ has only
one center at $r=r_i$ ; $i=1,2,3$. Thus, each $\Delta_i$ in
(\ref{721}) - (\ref{233}) has only single poles, which can be canceled
against each other. Eg.\ in $\Delta_1$ the poles from $K^2 K^3$ at
$r=r_2$ and at $r=r_3$ are canceled if $L_1$ has two poles also
$r=r_2$ and $r=r_3$. The same can be done for $\Delta_2$ and
$\Delta_3$.  In total these are six equations and in addition, there
are three equations coming from the $\omega_5$ constraint in
(\ref{726}), because this equation has poles at each center. So in
total, we have now 9 constraints. In general, we do not expect to
solve these constraints just by fixing the constant parts of the
harmonic functions (there are too few of them) and therefore, one may
have to impose constraints on the charges related to the separate
centers. But note, the total charges of the bound state measured at
infinity, can always be chosen freely. The positions are again fixed
by solving the integrability constraint. It would be very interesting
to work out this example in detail and see whether in fact all
continuous parameters are fixed at the end.  To make the story
complete, one has also to make sure that there are no closed timelike
curves, but also this issue can be addressed in four dimension; see
below. Further interesting questions are then: How can one understand
the entropy of this bound state? Can one break supersymmetry? Note, the
naive non-extreme black hole solution does not allow a multicenter
case, which is important in this setup.


\section{Relating four- and five-dimensional BPS solutions}


The whole class of solutions that we just summarize, can be mapped to
the known stationary solutions in four dimensions \cite{9705169}. This
is a one-to-one map, without any smearing or any other approximation
and different aspects of this relation are explored in
\cite{0408122,0503217,0506251}.  It is an important relationship,
because it opens the possibility to understand corrections to
(classical) 5-dim.\ supergravity. These are for example, higher
curvature or general higher derivative corrections or, since the
(asymptotic) 5-dim.\ space time has a compact circle (ie.\ $R_4\times
S_1$) one can expect even instanton corrections (as the one discussed
in \cite{9704095}) to 5-dim.\ solution, which are of course suppressed by
the radius of the $S_1$.

It is important to realize that this mapping relates localized solutions,
there is no smearing procedure along the $5th$ coordinate. To understand it,
consider the single center Reissner-Nordstrom black hole, which reads
in 5-dimensions
\be719
ds^2 = -\frac{1}{H^2} dt^2 + H \, [ d\rho^2 + \frac{\rho^2}{4}
(\sigma_1^2 + \sigma_2^2 + \sigma_3^2)]
\ , \quad A = \frac{dt}{H}  \ , \quad H = 1 + \frac{\rho_0^2}{\rho^2}
\ee
and becomes after the transformation $\rho^2 = 4 R\, r$
\be819
ds^2 = -\frac{1}{H} dt^2 + H\, \Big[ \frac{1}{N} R^2 \sigma_3^2
+ N (dr^2 + r^2[\sigma_1^2 + \sigma_2^2])\Big] \quad , \qquad
N = \frac{R}{r} \ .
\ee
The reduction along $x^5 = R \psi$ ($\sigma_3 = d\psi +
\cos^2\theta d\varphi$) yields
\be910
ds_4^2 = e^{2U} dt^2 + e^{-2U} [ dr^2 + r^2 d\Omega_2]
\quad , \qquad
e^{-2U} = \sqrt{N H^3} \ .
\ee
If one identifies $N$ and $H$, this is exactly the 4-dim.\
Reissner-Nordstrom black hole and note, the 5-dim.\ black hole was
localized in all four spacial directions!  A crucial point for the
mapping is that the entropy of both black holes should be equal.
In fact, one finds
\be512
S_5 = \frac{A_3}{4 G_5} = 2 \pi \sqrt{q^3} \equiv
\frac{A_2}{4 G_4} = S_4
\ee
where $A_{3/2}$ are the areas of the corresponding 3- and 2-dim.\ black hole
horizons. There is straightforward generalization: $N \rightarrow 1 +
\frac{pR}{r}$ so that $S_5 = S_4 = 2 \pi \sqrt{p q^3}$ and if the NUT charge
is $p=1$, the asymptotic space is effectively 4-dimensional ($S_1 \times
R_3$), but as $r \rightarrow 0 $ the geometry is equivalent to the
5-dimensional metric (\ref{819}) and therefore, by varying $R$, the geometry
interpolates between the 4- and 5-dimensional case \cite{0503217}.

It is now straightforward to include additional vector multiplets in
five dimensions and the resulting 4-dimensional geometry is given by
the stationary solutions discussed in \cite{9705169}.  All couplings
and fields of the 4-dim.\ Lagrangians can be expressed in terms of the
symplectic section $(Y^I , F_I)$ [$F_I = \frac{d}{dY^A} F(Y)$], which
has been rescaled by the central charge \cite{9610105}.  In order to
solve the gauge field equation and Bianchi identities one introduces
a set of harmonic functions
\be789
(H^I ; H_I) \quad = \quad (N, K^A ; M , L_A)
\ee
and supersymmetry requires that the the section obeys the attractor
equations
\be704
2 \,{\rm Im} Y^I = H^I \quad , \qquad 2\, {\rm Im} F_I = H_I \ .
\ee
The solution can then be written as
\bea221
Y^0 &=& \frac{1}{2} (\phi^0  + i N) \quad  , \qquad Y^A =
-\frac{|Y^0|}{\sqrt{N}} \, x^A + \frac{Y^0}{N} \, K^A
\ , \\ 
 \phi^0 &=& e^{2U} [ N^2 M + N L_A K^A + \frac{1 }{3} C_{ABC}K^AK^BK^C ]
\eea
where the variables $x^A$ are solutions of the set quadratic equations
\be765
\frac{1}{2} C_{ABC} x^B x^C = L_A + \frac{1}{2} \frac{C_{ABC} K^B K^C}{N}
\equiv \Delta_A \ .
\ee
The 4-dim.\  metric can then be written as
\bea625
ds^2 &=& - e^{2U} (dt + \vec \omega)^2 + e^{-2U} d\vec x^2 \ , \\
e^{-4U} &=& \frac{4}{9} N (\Delta_A x^A)^2 - \Big[ MN + L_A K^A +\frac{1}{3}
\frac{C_{ABC} K^A K^B K^C}{N} \Big]^2 \ ,  \label{299} \\
{^{\star_3}d}\vec \omega &=& NdM -MdN +K^A dL_A - L_A dK^A
\eea
and the integrability of the last equation becomes \cite{0005049}
\be413
0= N \Delta M -M \Delta N + K^A \Delta L_A - L_A \Delta K^A \ .
\ee
There is now the following dictionary between the 4- and 5-dim.\ quantities
\cite{0506251}
\be231
Y^A_{5d} = 2^{1/3} \, x^A \quad , \qquad
f^{-3/2} = \frac{2}{3} \, \Delta_A x^A \quad , \qquad
\omega_5 = R\, \frac{e^{-2U} \phi^0 }{N^2} \ .
\ee
It is straightforward to verify that the 4-dim.\ solution is regular
at $N=0$ (all terms $\sim 1/N$ or $\sim 1/N^2$ cancel) and therefore
also the 5-dim.\ solution has to be regular.  The effective angular
momentum in five dimensions is related to the combination of harmonic
functions $J \leftrightarrow M + \frac{L_A K^A}{N} +\frac{1}{3}
\frac{C_{ABC} K^A K^B K^C}{N^2}$ and the condition $e^{-2U}>0$ sets an
upper bound on the angular momentum beyond which the 5-dim.\
solution develops closed timelike curves whereas the 4-dim.\ becomes
singular. For the multi-center case, the constraint (\ref{413}) can be
interpreted that the charge vectors $(p^I , q_I)$ of the different
centers as well as the symplectic vector $(H^I, H_I)_{r= \infty}$ have
to be mutual local.

This was the general solution, but there are a number of interesting examples:
(i) The simplest one are the well-known single-center black holes with $H^I =
h^I + \frac{p^I R}{r}$, $H_I = h_I + \frac{q_I R}{r}$, which describes in five
dimensions a rotating black hole in a Taub-NUT space (the angular momentum as
well as the NUT charge become an electric and magnetic charge in four
dimensions). (ii) Black rings are mapped on a 2-center solution in four
dimensions; one center is regular and has no NUT charge, but the other carries
the NUT charge and is singular. (iii) The G\"odel type solution is obtained if
$M = g z$ (ie.\ a linear function in one coordinate, $g=const.$), $N= R/r$,
$L=const.$ and $K=0$. Since $M$ is linearly growing function, the solution in
four dimensions develops a curvature singularity at finite radius: $r=r_c$
where $e^{-2U} = 0$. As for the over-rotating BMPV black hole, this
singularity is related the vanishing circle of the $5th$ direction.  We should
also mention, that the 4-dim.\ solution can develop closed timelike curves if
$\det(-e^{2U} \omega_i \omega_j + e^{-2U} \delta_{ij}) < 0$ ($i,j =
1,2,3$). This constraints as well as the absence of Dirac-Misner strings have
to be ensured for consistency.  (iv) It is straightforward to combine the
different solutions by adding the appropriate harmonic functions. Black holes
in a G\"odel space time corresponds to adding a linear function to at least
one of the harmonic functions \cite{0401239}. Although this correction drops
out on the black hole horizon, it will always yield a singularity at some
finite radius.  Within supergravity, the only way to avoid this singularity is
to cut-off the region in space time by introducing a domain wall by $gz
\rightarrow g (1- |z -z_0|)$, where $z_0$ should be chosen appropriately. But
as always in such scenarios, one has to address the question, what happens if
the wall moves close to the singularity?  (v) We should also note, that the
reduction of the 5-dim.\ solution over the $S^2$ instead of the non-trivial
$S_1$ circle yields the BTZ black hole in three dimension and again the
entropy agrees with the one in five and four dimensions.


\begin{acknowledgement}

I would like to thank Anna, Max and Gabi as well as all of my collaborators
for the encouraging support over many years.  ``Oh, and in case I don't see
ya. Good afternoon, good evening, and good night''.
\end{acknowledgement}



\providecommand{\href}[2]{#2}\begingroup\raggedright\endgroup


\end{document}